\pdfoutput=1
\documentclass[10pt]{article}
\usepackage{wrapfig,cancel,tikz,soul,ulem,framed,caption, todonotes}
\definecolor{shadecolor}{rgb}{0.95, 0.95, 0.86}

\usetikzlibrary{decorations.markings,arrows}

\usepackage{color}
\def \Etavertex{ \eta_{_{\mathfrak V}}}
\def \Thetavertex{ \theta_{_{\mathfrak V}}}
\usepackage{graphicx}
\def\g{\gamma}
\usepackage[colorlinks=true, pdfstartview=FitV, linkcolor=blue, citecolor=blue,urlcolor=blue]{hyperref}

\def \QED {\hfill $\blacksquare$\par \vskip 3pt}
\def\Xint#1{\mathchoice
{\XXint\displaystyle\textstyle{#1}}%
{\XXint\textstyle\scriptstyle{#1}}%
{\XXint\scriptstyle\scriptscriptstyle{#1}}%
{\XXint\scriptscriptstyle\scriptscriptstyle{#1}}%
\!\int}
\def\XXint#1#2#3{{\setbox0=\hbox{$#1{#2#3}{\int}$ }
\vcenter{\hbox{$#2#3$ }}\kern-.6\wd0}}

\def\slint{\Xint-}

\def\green#1{\textcolor[rgb]{0.2, 0.5,  0} {#1}}
\def\cprime{$'$}
\def \G{\Gamma}

\usepackage{amsbsy}
\usepackage{amssymb}
\newtheorem{problem}{Problem}[section]
\def\D{\mathbb D}

\usepackage{verbatim}
\def\tr {\mathrm {Tr}}

\makeatletter
\@addtoreset{equation}{section}
\makeatother

\newtheorem{theorem}{Theorem}[section]
\newtheorem{example}{Example}[section]
\newtheorem{exercise}{Exercise}[section]

\newtheorem{lemma}{Lemma}[section]
\newtheorem{remark}{Remark}[section]

\newtheorem{proposition}{Proposition}[section]
\newtheorem{corollary}{Corollary}[section]
\newtheorem{definition}{Definition}[section]
\def\le{\left}
\def \eqref#1{(\ref{#1})}
\def\ri{\right}
\def\ds{\displaystyle}

\def\br{\begin{remark}}
\def\er{\end{remark}}
\def\bt{\begin{theorem}}
\def\et{\end{theorem}}
\def\bc{\begin{corollary}}
\def\ec{\end{corollary}}
\def\bx{\begin{example}\small}
\def\ex{\end{example}}
\def\bxr{\begin{exercise}\small}
\def\exr{\end{exercise}}
\def\bl{\begin{lemma}}
\def\el{\end{lemma}}

\def\ddz{ \frac{{\rm d }z} {2i\pi}}
\def\ddw{ \frac{{\rm d }w} {2i\pi}}
\def\bd{\begin{definition}}
\def\ed{\end{definition}}
\def\bp{\begin{proposition}}
\def\ep{\end{proposition}}
\def\be{\begin{eqnarray}}
\def\ee{\end{eqnarray}}
\def\&{\hspace{-15pt}&}
\def\bea{\begin{eqnarray}}
\def\eea{\end{eqnarray}}
\def\beas{\begin{eqnarray*}}
\def\eeas{\end{eqnarray*}}

\def \pa{\partial}
\def\C{{\mathbb C}}

\def\R{{\mathbb R}}

\def\d{\,\mathrm d}

\def\l{\lambda}

\def\1{{\bf 1}}
\def\wt{\widetilde}

\date{}
\begin{document}
\baselineskip 15pt plus 1pt minus 1pt

\vspace{0.2cm}
\begin{center}
\begin{Large}
\fontfamily{cmss}
\fontsize{17pt}{27pt}
\selectfont
\textbf{{\sc Corrigendum:}
The dependence on the monodromy data  of the isomonodromic tau function}
\end{Large}\\
\bigskip
\begin{large} {M.
Bertola}$^{\dagger,\ddagger\sharp}$\footnote{Work supported in part by the Natural
    Sciences and Engineering Research Council of Canada
(NSERC).}\footnote{Marco.Bertola@\{concordia.ca, sissa.it\}.}
\end{large}
\\
\bigskip
\begin{small}
$^{\dagger}$ {\it Department of Mathematics and
Statistics, Concordia University\\ 1455 de Maisonneuve W., Montr\'eal, Qu\'ebec,
Canada H3G 1M8} \\
$^{\ddagger}$ {\it SISSA/ISAS, via Bonomea 265, 34135, Trieste, Italy}\\
$^{\sharp}$ {\it Centre de recherches math\'ematiques,\  Universit\'e\ de
Montr\'eal } \\
\end{small}
\bigskip
{\bf Abstract.}
\end{center}
The note corrects the aforementioned paper \cite{BertolaIsoTau}. The consequences of the correction are traced and the examples updated.

\section{Introduction}
With this note I wish to correct two mistakes of loc. cit.; one of them (and its consequences) is of trivial nature and it is quickly disposed of. 
The second is significant, but not fatal, and requires rather extensive additional material, which is the reason of the relatively large size of this erratum.

I also wish to  thank  Oleg Lisovyy  for pointing out certain inconsistencies  that convinced me that indeed there was a mistake in need of fixing. The mistake was observed during the preparation of \cite{ItsProkhorov} and \cite{ItsProkLis} because it was not consistent with their explicit computations: this prompted one of the authors to contact me.

The correction only affects certain types of setup; if the contours supporting the jumps of the Riemann--Hilbert Problem do not intersect, then the result is correct as it stands. 
In the subsequent works of my collaborators and myself  only this type of problems were actually considered and therefore the following papers are  essentially unaffected (although any  reporting of the original formula is not correct in the stated generality):
\cite{bertolacafasso5, BertolaCafasso4, BertolaCafasso3, BertolaCafasso2, BertolaCafasso1}.
\section{Minimal setup}
The original setup requires to consider a Riemann--Hilbert Problem (RHP) with the following 
data  (here reformulated with greater detail than in the original paper)
\paragraph{The Riemann--Hilbert data}
\begin{enumerate}
\item a finite collection of smooth oriented arcs $\gamma_\nu,\ \nu=1\dots K$,
possibly meeting at a finite number of points but always in non-tangential way. We denote  collectively these arcs by the symbol $\ds \Sigma \gamma = \bigcup \gamma_\nu$. 
\item a collection of  $r\times r$ matrices $M_\nu(z)$, each of which analytic at each interior point of its corresponding arc $\gamma_\nu$.    We will denote collectively by $M(z)$ the matrix defined on $\Sigma \gamma$ that coincides with $M_\nu(z)$ on $\gamma_\nu$,  
\bea
M: \Sigma\gamma&\to& SL_r(\C)\cr
z & \mapsto & \sum_{\nu} M_\nu(z) \chi_{\gamma_\nu}(z)
\eea
where, for a set $S$,  $\chi_S$ denotes its indicator function.
\item  At each point $c$ where several arcs meet, denoting by  $\gamma_1,\dots, \gamma_\ell$ the arcs entering  a suitably small disk at $c$, we require that the arcs approach $c$ along distinct, well-defined directions and we impose that the jump matrices along its corresponding arc either \begin{itemize}
\item  admit local analytic extension within said disk. 
In this case,  if we denote by $\gamma_1,\dots, \gamma_n$ the contours incident at  $c$, oriented outwards, and labelled counterclockwise,  and $M_\ell (z) = M(z)\bigg|_{z\in \gamma_\ell}$, we require that  the aforementioned analytic extensions satisfy
\be
M_1(z)\cdot M_2(z) \cdots M_n(z) \equiv \1, 
\label{localnomonodromy}
\ee
and this equality holds 
(locally) identically also with respect to the deformation parameters. Such an intersection point will be referred to as ``essential" later on, for lack of better word.
 
\item   tend to the identity matrix as $\mathcal O((z-c)^\infty)$ (faster than any power) in an open sector containing the direction of approach (this applies also to any jump matrix on contours extending to infinity, where $(z-c)$ is replaced by $1/z$) and admit analytic continuation on the universal cover of the punctured disk around $c$. Such an intersection point will be referred to as ``inessential".
\end{itemize}
\end{enumerate}

\begin{problem}[RHP]
\label{probRH}
Find a holomorphic matrix $\Gamma:\C\setminus \Sigma\gamma\to GL_n(\C)$ such that 
\begin{itemize}
\item $\Gamma_+(z) = \Gamma_-(z) M(z)$ $z\in \Sigma\gamma$;
\item $\Gamma(z),\ \Gamma^{-1}(z) $ are uniformly bounded in $\C$;
\item $\Gamma(z_0)=\1$
\end{itemize}
\end{problem}
Assuming that the solution exists for given initial data, \cite{BertolaIsoTau} considered the  {\it deformations} of the jump matrices (respecting the conditions listed above). 
\br
The conditions  on the jump  matrices laid out above ensure 
that the solution $\Gamma(z)$ admits analytic continuation in a neighbourhood of the intersection point $c$,  or at least in the  open sector around the direction of approach mentioned above. In this latter case the decay condition guarantees 
 that the solution admits an asymptotic expansion near $c$ in the same sector, and that the expansion coefficients do not depend on the sector. The conditions are modelled upon the case of RHPs associated to rational ODEs in the complex plane. 
\er
\subsection{Corrections}
 The overall minus sign in (2.7) of Def. 2.1 \cite{BertolaIsoTau} should be removed. While this is a definition, the purpose was to extend the Jimbo-Miwa-Ueno definition, and the correct sign should have been the opposite one. For convenience, here is the corrected definition. Of course the sign should be changed also in the subsequent formul\ae. 
\bd[Def 2.1 in \cite{BertolaIsoTau}]
\label{defomega}
Let $\pa$ denote the derivative w.r.t. one of the parameters $s$ and assume that the Riemann--Hilbert Problem \ref{probRH} admits a solution in an open subset of the deformation--parameter space.\footnote{The small--norm theorem for Riemann--Hilbert problems implies that if a RHP is solvable, then any sufficiently small deformation (in $L^2$ and $L^\infty$ norms) of the jump matrices leads to a solvable RHP. With our assumptions on the $s$--dependence of the jump matrices this implies that the subset of solvable RHP is an open set (if non-empty).} Then we define Malgrange's form $\omega_{_M}$ 
\bea
\omega_{_M}(\pa) = \omega_{_{M}}(\pa; [\Gamma]) :=   \int_{\Sigma \gamma} \tr \bigg(\Gamma^{-1}_-(z) \Gamma_-'(z)  \Xi_\pa (z)\bigg) \ddz \label{omegaM}\qquad \qquad
\Xi_\pa(z):= \pa M(z) M^{-1}(z)\ .
\eea
\ed

\paragraph{New notation}
In order to deal more expeditiously with the correction we shall also use the matrix-valued forms (Maurer-Cartan like) $\Xi(z):= \delta M(z) M^{-1}(z)$, where $\delta$ shall denote henceforth the exterior derivative with respect to the deformation parameters $\vec t$ (not to confuse it with  $\d z$ of the spectral variable). We shall also retain the notation $\Xi_\pa$ for the contraction of said form with a vector field $\pa$.

Proposition 2.1 in \cite{BertolaIsoTau} offers an incomplete formula for the exterior derivative of $\omega_{_{M}}$ and we correct it now. The additional term in the following Theorem is present only when there are points of $\Sigma\gamma$ with  several incident arcs; we call this  the ``set of vertices'' of $\Sigma\gamma$ and denote it by $\mathfrak V$. 
If $\Sigma\gamma$ consists in the union of smooth  disjoint arcs, or all the jump matrices tend to the identity at all the vertices, then the original statement stands correct.

\begin{shaded}
\bt[Replaces Prop. 2.1 of \cite{BertolaIsoTau}]
\label{thmreplace}
Denote by $\mathfrak V\ni v$ the vertices of the graph $\Sigma\gamma$; let $\mathcal E_v = \bigcup_{j=1}^{n_v} \gamma_j^{(v)}$ be the set of arcs incident to $v$, oriented outwards and enumerated counterclockwise.
Then exterior derivative of $\omega_{_{M}}$ is 
\be
\delta\omega_{_{M}} =&\& - \frac 1 2 \int_{\Sigma\gamma}  \ddz \tr\bigg(
\frac {\d}{\d z} \Xi (z) \wedge \Xi(w)
\bigg)\bigg|_{w=z}
+\Etavertex
\label{deltaomega24}
\ee
with 
\be
\Etavertex :=  \frac {\mathbf -1}{4i\pi}\sum_{v\in \mathfrak V} \sum_{\ell=2}^{n_v} \sum_{m=1}^{\ell-1} \tr \bigg(
M_{[1:m-1]}^{(v)} \Xi^{(v)}_{m}M^{(v)}_{[m:\ell-1]}\wedge \Xi^{(v)}_{\ell} M^{(v)}_{[\ell:n_v]} 
\bigg)
\label{corr21}
\ee
\be
\Xi^{(v)}_\ell  = \lim_{z\to v} \delta M_{\ell}^{(v)} (M_{\ell}^{(v)})^{-1}\bigg|_{z\in \gamma_{\ell}^{(v)}}\qquad 
M_{\ell}^{(v)} = \lim_{z\to v\atop z\in \gamma_{\ell}^{(v)}} M^{\epsilon_\ell} (z).
\ee
where the power $\epsilon_\ell= 1$ if the contour $\gamma_\ell^{(v)}$ is oriented away from $v$ and $\epsilon_\ell= -1$ if oriented towards. Here the subscript $_{[m:\ell-1]}$ is a shorthand to signify the product of the corresponding matrices over the range of indices $m,m+1,\dots, \ell-1$. 
\et
\end{shaded}
The complete proof is reported in Section \ref{proof}.
The "modified Malgrange form" $\Omega$ (Def. 2.2 in \cite{BertolaIsoTau}) is (with the corrected sign)
\bd
\label{defOmega}
The modified Malgrange differential is defined as $\Omega:= \omega_{_M} +\vartheta$ with 
\be
\vartheta(\pa):= \frac 1 2 \int_{\Sigma\gamma} \tr \le(M'(z)M^{-1}(z) \pa M(z) M^{-1}(z)\ri)\ddz\label{recurv}
\ee
Equivalently (see (2.32) in \cite{BertolaIsoTau})
\be
\Omega(\pa;[\Gamma]) =\frac 1 2 \int_{\Sigma\gamma} \tr \le( 
 \Gamma_-^{-1} (z)\Gamma_-' (z)\pa M(z) M^{-1}(z) +\Gamma_+^{-1}(z) \Gamma_+'(z) M^{-1}(z) \pa M (z)
\ri)\ddz
\ee
\ed
Consequent to the correction of Prop. 2.1, the ancillary result below (Prop. 2.2 in \cite{BertolaIsoTau}) is also similarly modified
\bp
\label{propcurv}
The curvature of the modified Malgrange form is 
\be
\delta \Omega =- \frac 1 2 \int_{\Sigma \gamma} \tr \bigg(M'(z) M^{-1}(z) \Xi(z) \wedge \Xi (z)\bigg)\ddz + \Etavertex
\ee
\ep

\subsection{Rational differential equations (amended)}
In the setup of Sec. 2.2, the term $\Etavertex$ is closed and admits a potential $\Thetavertex$; consequently  in Sec. 2.2 of \cite{BertolaIsoTau}  the title should read: 
"Submanifolds of $\mathcal G$ where $ \delta \Omega -\Etavertex =0$".
Sections 2.3, 2.4 are unaffected. 

The main application of the original paper was to Riemann--Hilbert problems related to the setting of \cite{JMU1}, i.e. the generalized monodromy data associated to a (generic) rational connection on $\C \mathbb P^1$. 

The statement that $\Omega$ is a closed one-form is incorrect in the stated generality and needs to be corrected.

To explain the necessary modifications we keep the same setup  of Sections 3,4,5 (and  Fig. 5, 6 of \cite{BertolaIsoTau}). 
\begin{shaded}\bt[Replaces Thm. 5.1 in \cite{BertolaIsoTau}]
\label{thmcorr}
There exists a locally defined one form $\Thetavertex = \Thetavertex(\vec L, \vec S, \vec C) $ on the manifold of generalized monodromy data (independent of the Birkhoff invariants and the position of the poles) such that the Jimbo-Miwa-Ueno tau function satisfies
\be
\label{taucorr}
\delta \ln \tau(\vec T, \vec a, \vec L, \vec S, \vec C) =  \omega_{_M} - \Thetavertex,
\ee
where $\delta \Thetavertex = \Etavertex$ in \eqref{corr21}. 
This function is defined up to nonzero multiplicative constant and it vanishes precisely when the Riemann--Hilbert problem is not solvable, namely, on the Malgrange Theta-divisor.
\et
\end{shaded}
\br
Note that the theorem is now stated directly in terms of $\omega_{_{M}}$ rather than the "modified" form $\Omega$ used in the original paper.  The two forms differ by an explicit one form so there is little simplification in choosing one over the other, since neither is closed by itself  in the relevant case.
\er
\br
Since $\Thetavertex$ is only locally defined on the monodromy manifold, the formula \eqref{taucorr} allows to identify $\tau$ as a section of a line bundle on said manifold. The transition functions are given by $\delta \ln g = \wt \Thetavertex-\Thetavertex$ on the overlap of two open charts. 
This observation, which stems from the correction term in Thm. \ref{thmreplace} seems to be of interest in applications that  are arising from recent works \cite{ItsProkhorov} and deserves further study.
\er
Rather than chasing a complete generality we illustrate the statement in several significant cases. 
\subsection{Example 0: Scalar Fuchsian case}
\begin{figure}
\begin{center}
\begin{tikzpicture}[scale=01.5]

\draw [fill=gray!10!white] (  0  :2) circle [radius =0.5];
\draw [fill] (  0  :2) circle [radius =0.02];
\draw [postaction={decorate,decoration={markings,mark=at position 0.5 with {\arrow[black,line width=1.5pt]{<}}}}](0:0)--(  0  : 1.5);
\node [above] at (  0:2) {$a_1$};
\node at (  5  :1.4) {$\beta_1$};
\node at (  -10  :2.0) {$\D_1$};

\draw [fill=gray!10!white] (  50  :2) circle [radius =0.5];
\draw [fill] (  50  :2) circle [radius =0.02];
\draw [postaction={decorate,decoration={markings,mark=at position 0.5 with {\arrow[black,line width=1.5pt]{<}}}}](0:0)--(  50  : 1.5);
\node [above] at (  50  :2) {$a_2$};
\node at (  56  :1.4) {$\beta_2$};
\node at (  41  :1.9) {$\D_2$};

\draw [fill] (  90  :1.5) circle [radius =0.02];
\draw [fill] (  130  :1.5) circle [radius =0.02];
\draw [fill] (  180  :1.5) circle [radius =0.02];
\node at (-0.1,0.1) {$z_0$};
\node at (-15:0.8) {$\gamma_1^{0}$};

\draw [fill=gray!10!white] (  200  :2) circle [radius =0.5];
\draw [fill] (  200  :2) circle [radius =0.02];
\draw [postaction={decorate,decoration={markings,mark=at position 0.5 with {\arrow[black,line width=1.5pt]{<}}}}](0:0)--(  200  : 1.5);
\node [below] at (  200  :2) {$a_n$};
\node at (  207  :1.3) {$\beta_n$};
\node at (  192 :1.9) {$\D_n$};
\end{tikzpicture}
\end{center}
\caption{The arrangement of disks for the scalar case.}
\label{scalar}
\end{figure}
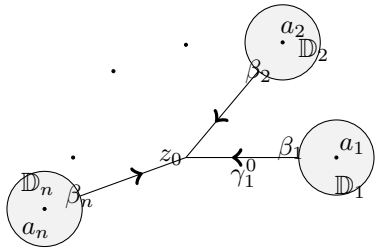
Consider the scalar RHP (see Fig. \ref{scalar})
\bea
\Gamma (z) = \le\{
\begin{array}{cc}
\prod_{j=1}^n \le(z-a_j\ri)^{\theta_j}  & \C \setminus \bigcup \D_j\\
\prod_{j=1, j\neq k}^n \le(z-a_j\ri)^{\theta_j}  & z\in \D_k
\end{array}
\ri.\ ,\cr
 \sum \theta_j =0.
\eea
\be
\Gamma_-^{-1} \Gamma_+  = 
\le\{
\begin{array}{cc}
(z-a_k)^{-\theta_k}  & z\in \gamma_k = \pa \D_k\\
{\rm e}^{-2i\pi \theta_k} & z\in \gamma_k^0
\end{array}
\ri.
\ee

\noindent where $\gamma^0_k$ is a contour $[\beta_k, z_0]$, with $\beta_k$ chosen and fixed on $\gamma_k$ and $z_0$ a fixed basepoint outside, chosen in such a way that no two  points $a_j$ are not on the same ray from $z_0$.  The Malgrange one-form is 
\bea
\omega_{_M} &\&= \int \Gamma_{-}^{-1} \Gamma_-' \delta M M^{-1}\ddz
=\\
&\&=\sum_{j=1}^n \int \frac {\theta_j}{z-a_j} \sum_{k}\le( \le(\frac{ \theta_k \d a_k }{z-a_k} - \ln (z-a_k)\d \theta_k \ri)\chi_{_{\pa\D_k}} - {2i\pi }\d \theta_k \chi_{_{\gamma^0_k}}\ri) \ddz = \hbox{(contour deformation)}
\cr
&\&= \sum_{j=1}^{n}\le[ \sum_{k\neq j}\le( \frac { \theta_j \theta_k \d a_k}{a_k- a_j}  -\theta_j \d \theta_k \int_{a_k}^{z_0} 
\frac{\d z }{z-a_j}   \ri)-
\frac  1 {4i\pi} \theta_j\d\theta_j {\ln^2}  (z-a_j)\bigg|_{\beta_j}^{\beta_j + \gamma_j}  - \theta_j \d \theta_j \int_{\beta_j}^{z_0} \frac {\d z}{(z-a_j)}\ri] 
\nonumber
\cr
&\&= \sum_{j=1}^{n}\Bigg[ \sum_{k\neq j}\le( \frac { \theta_j \theta_k \d a_k}{a_k- a_j}  -\theta_j \d \theta_k \big(\ln(z_0-a_j) - \ln (a_k-a_j)\big) \ri)+
\cr
&\&\qquad -
\frac  1 {4i\pi} \theta_j\d\theta_j (4i\pi \ln(\beta_j - a_j) + (2i\pi)^2) -  \theta_j \d \theta_j (\ln(z_0-a_j) - \ln (\beta_j - a_j))\Bigg] 
\nonumber
\eea
Here all logarithms are principal; the term involving $z_0$ drop out because $\sum \d \theta_k =0$, as well as the dependence on $\beta_j$. We are left with 
\be
\omega_{_{M}} =  \sum_{j=1}^{n}\Bigg[ \sum_{k\neq j}\le( \frac { \theta_j \theta_k \d a_k}{a_k- a_j}  +\theta_j \d \theta_k \ln (a_k-a_j)\ri)
 -
i\pi \theta_j\d\theta_j\Bigg] 
\ee
Here the logarithms are all principal. 
The exterior derivative of the above expression is 
\be
\delta \omega_{_{M}} = \Etavertex = \sum_j\sum_{k\neq  j} \d \theta_j\wedge \d \theta_k(\ln (a_k-a_j)- \ln(a_j-a_k)) ={ i\pi\sum_j\sum_{k<  j} \d \theta_j\wedge \d \theta_k = i\pi \d\le( \sum_{j}\sum_{k<j} \theta_j \d\theta_k\ri)
}
\ee
In this case the Tau function is explicit 
\be
\delta \ln \tau = \omega_{_{M}} +i\pi \sum_{k<\ell} \theta_k \d \theta_\ell \ ,\ \ \ 
\tau (\vec a, \vec \theta) = \prod_{\ell=1}^n\prod_{k<\ell} (a_k-a_\ell)^{\theta_k\theta_\ell} \prod_{k=1}^n {\rm e}^{-\frac {i\pi}2 \theta_k^2} 
\ee
To be noted, there is an ambiguity in the above writing because of the determinations of the logarithm; the ambiguity is what defines the line bundle of which $\tau$ is a section.

\subsection{Fuchsian singularities with nontrivial  monodromy (e.g. Painlev\'e\ VI)}

\begin{figure}
\begin{center}
\begin{tikzpicture}[scale=01.8]
\begin{scope}[shift={(15:2)}]
\draw[postaction={decorate,decoration={markings,mark=at position 0.9 with {\arrow[line width=1.5pt]{>}}}}]
circle [radius=1];
\draw[thick, postaction={decorate,decoration={markings,mark=at position 0.1 with {\arrow[line width=1.5pt]{>}}}}]
circle [radius=1];

\draw[green!60!black, thick,postaction={decorate,decoration={markings,mark=at position 0.9 with {\arrow[line width=1.5pt]{>}}}}] circle [radius=0.5];

\node [above] at (90:0.5){\resizebox{1.2cm}{!}{$(z-a_1)^{-L_1}$}};
\node [below]at (-90:1){$C_1^{-1}$};
\end{scope}

\draw[postaction={decorate,decoration={{markings,mark=at position 0.65 with {\arrow[black,line width=1.5pt]{>}}}} }]
(0,0)--node[above]{{$\mathcal M_1$}}(15:1);
\draw[postaction={decorate,decoration={{markings,mark=at position 0.75 with {\arrow[black,line width=1.5pt]{>}}}} }] 
(15:1)-- node[above]{\resizebox{0.7cm}{!}{${\rm e}^{2i\pi L_1}$}}(15: 1.5);

\draw[postaction={decorate,decoration={{markings,mark=at position 0.65 with {\arrow[black,line width=1.5pt]{>}}}} }]
(0,0)--node[above]{{$\mathcal M_{_K}$}}(180:1);
\draw[postaction={decorate,decoration={{markings,mark=at position 0.75 with {\arrow[black,line width=1.5pt]{>}}}} }]
(180:1)--node[above]{\resizebox{0.7cm}{!}{${\rm e}^{2i\pi L_K}$}}(180: 1.5);

\node [below] at (0,0) {$z_0$};
\foreach \x in {-1,-0.5,0.5,0.8}
 \draw [fill] (\x, 1)  circle [radius=0.01];
 
\node [right] at (15:1.5)  {$\beta_1$} ;
\draw (15:1.5)  circle[radius=0.02];
\node [left] at (184:1.4)  {$\beta_{_K}$} ;
\draw (180:1.5)  circle[radius=0.02];

\begin{scope}[shift={(180:2)}]
\draw[postaction={decorate,decoration={markings,mark=at position 0.9 with {\arrow[line width=1.5pt]{>}}}}]
circle [radius=1];
\draw[thick, postaction={decorate,decoration={markings,mark=at position 0.1 with {\arrow[line width=1.5pt]{>}}}}]
circle [radius=1];
\draw[green!60!black, thick,postaction={decorate,decoration={markings,mark=at position 0.9 with {\arrow[line width=1.5pt]{>}}}}] circle [radius=0.5];
\node [above] at (90:0.5){\resizebox{1.2cm}{!}{$(z-a_{_K})^{-L_{_K}}$}};
\node [below]at (-90:1){$C_{_K}^{-1}$};
\end{scope}

\end{tikzpicture}

\end{center}
\caption{
The arrangement of jumps for a generic Fuchsian system.  }
\label{Schles2RHP}
\end{figure}
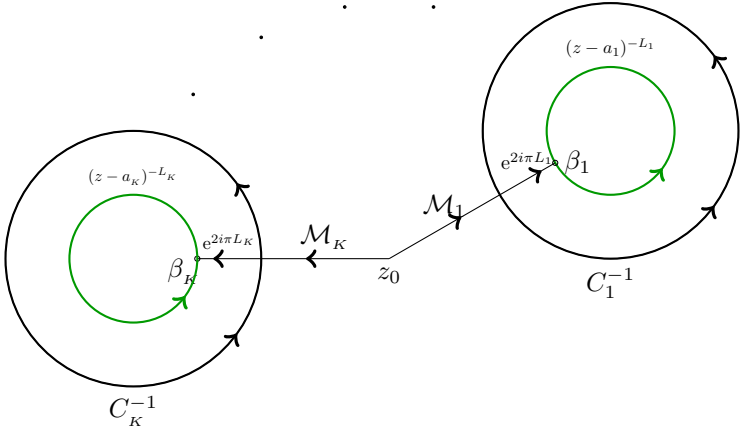 
Suppose that the RHP corresponds to the solution of a generic Fuchsian ODE with simple poles at $a_1,\dots, a_K$ of the form 
\be
\Psi'(z) = \sum_{j=1}^K\frac{A_j}{z-a_j} \Psi(z)\  ,\ \ \ A_j = O_j L_j O_j^{-1}, \ \ \ L_j = \hbox{ diagonal }.
\ee

We set $\Lambda_j = {\rm e}^{2i\pi L_j}$ (diagonal) and  the monodromy matrices are $\mathcal M_j := C_j^{-1} \Lambda_j C_j$. 
The enumeration is counterclockwise from the basepoint $z_0$ as indicated in Fig. \ref{Schles2RHP}.
We have the condition 
\be
\mathcal M_1 \cdots \mathcal M_K = \1
\label{chvar}
\ee

Then a direct computation using Thm. \ref{thmreplace}
yields 
\be
\delta \omega_{_{M}}  =&\& \frac {-1}{4i\pi}
\Bigg(
\sum_{\ell=2}^K \sum_{1\leq \ell< k} \tr \bigg(  \mathcal M_{{[1:\ell-1]}} \delta \mathcal M_{\ell} \mathcal M_{{[\ell+1: k-1]}}\wedge  \delta \mathcal M_{{k}}\mathcal M_{{[ k+1:K]}} \bigg)
+\cr
&\&+ \sum_{\ell=1}^K \tr \bigg(   \Lambda_\ell \delta C_{\ell}C_{\ell} ^{-1}\wedge  \Lambda_\ell^{-1} \delta C_\ell C_\ell ^{-1}+2 \Lambda_\ell ^{-1}  \delta \Lambda_\ell  \wedge \delta C_\ell C_\ell ^{-1}  \bigg) \Bigg)\label{dOmega}
\ee
As announced in Thm. \ref{thmcorr},  $\delta \omega_{_{M}} $ is {\it independent of the poles' positions}. It is a closed  two-form on the monodromy variety \eqref{chvar} itself.
It is not immediate to verify directly from the formula that the two--form  in  \eqref{dOmega} is actually closed, but it is a consequence of the Theorem \eqref{thmreplace}\footnote{The verification that the expression \eqref{dOmega} is indeed closed on the manifold \eqref{chvar} was performed also directly with the aid of a computer in small cases; a direct proof would be desirable.}.

On the other hand, since it is a closed two-form on the monodromy variety \eqref{chvar}, it follows that there is a locally defined one-form on the monodromy variety, which we shall denote $\Thetavertex$, such that $\delta \Thetavertex = \delta \omega_{_M}$.

We remark for the reader that 
the left hand side of \eqref{dOmega} is the result of a partial cancellation of terms between the two terms in \eqref{deltaomega24}.

\bx
The simplest example of Painlev\'e\ VI requires to describe explicitly the one-form $\Thetavertex$. 

We shall assume that the monodromies $\mathcal M_{1,..4}$ are non-resonant (i.e. the eigenvalues of $L_j$ do not differ by integers).
 This, however,  proves to be too complicated to handle explicitly in complete generality,  so we consider, by the way of example, the following particular submanifold of \eqref{chvar};
 \be
C_1  = \le[
\begin{array}{cc}
1& s_1\\
0&1
\end{array}
\ri],\ \ 
C_2  = \le[
\begin{array}{cc}
1& 0\\
s_2&1
\end{array}
\ri],\ \ 
C_3  = \le[
\begin{array}{cc}
1& s_3\\
0&1
\end{array}
\ri],\ \ 
C_4  = \le[
\begin{array}{cc}
1& 0\\
s_4&1
\end{array}
\ri]
\ee
We set $\Lambda_j = {\rm diag}(\l_j, \l_j^{-1})$; then we can solve the condition \eqref{chvar} for $\l_4, s_2, s_4 $ on a suitable open subset of the above submanifold of the monodromy variety

\be
 \bigg\{ &\&  \lambda_{{4}}=-{\frac {s_{{3}}\lambda_{{1}}\lambda_{{2}}
 \left( \lambda_{{3}}^2-1 \right) }{
\lambda_{{3}}s_{{1}} \left( \lambda_{{1}}^2-1 \right)   }},\ \ 
s_{{2}}={\frac {s_{{3}}{\lambda_{{1}}}^{2}{\lambda_{{
2}}}^{2}{\lambda_{{3}}}^{2}-s_{{3}}{\lambda_{{1}}}^{2}{\lambda_{{2}}}^
{2}+{\lambda_{{1}}}^{2}s_{{1}}-s_{{1}}}{s_{{3}}s_{{1}} \left( \lambda_
{{3}}^2-1 \right)   \left( \lambda_{{2}}^2-
1 \right)  \left( \lambda_{{1}}^2-1
 \right)  }},
 \cr
&\& s_{{4}}=-{\frac { \left( s
_{{3}}{\lambda_{{1}}}^{2}{\lambda_{{2}}}^{2}{\lambda_{{3}}}^{2}-s_{{3}
}{\lambda_{{1}}}^{2}{\lambda_{{2}}}^{2}+{\lambda_{{1}}}^{2}s_{{1}}-s_{
{1}} \right) {\lambda_{{3}}}^{2}}{ \left( {\lambda_{{1}}}^{2}\lambda_{
{3}}s_{{1}}-\lambda_{{3}}s_{{1}}+s_{{3}}\lambda_{{1}}\lambda_{{2}}-s_{
{3}}\lambda_{{1}}\lambda_{{2}}{\lambda_{{3}}}^{2} \right)  \left( {
\lambda_{{1}}}^{2}\lambda_{{3}}s_{{1}}-\lambda_{{3}}s_{{1}}+s_{{3}}
\lambda_{{1}}\lambda_{{2}}{\lambda_{{3}}}^{2}-s_{{3}}\lambda_{{1}}
\lambda_{{2}} \right) }} \bigg\} 
\ee
so that the monodromy  variety is now locally coordinatized $\l_1,\l_2,\l_3, s_1,s_3$. 
Then the two form $\delta \omega_{_{M}}$ is 
\be
\delta \omega_{_{M}} =
{\frac {( {\lambda_{{1}}}^{2}+1 )d\lambda_1\wedge d\lambda_2  }{2i\pi \lambda_{{2}} \left( {\lambda_{{1}}}^{2}-1 \right) \lambda_{{1}}}}
+
\frac {({\lambda_{{1}}}^{2}{\lambda_{{3}}}^{2}+{\lambda_{{3}}}^{2}+{\lambda_{{1}}}^{2}+1) d\lambda_1\wedge d\lambda_3
}{ 2i\pi \left( {\lambda_{{3}}}^{2}-1 \right) \lambda_{{3}}\lambda_{{1}} \left( {\lambda_{{1}}}^{2}-1 \right) }
-
{\frac { d\lambda_1\wedge ds_1}{2i\pi \lambda_{{1}}s_{{1}}}}
+
\frac {({\lambda_{{1}}}^{2}+1) d\lambda_1\wedge ds_3}{2i\pi \lambda_{{1}}s_{{3}} \left( {\lambda_{{1}}}^{2}-1 \right) }\,
+\cr
-
{\frac {({\lambda_{{3}}}^{2}+1) d\lambda_2\wedge d\lambda_3 }{2i\pi \lambda_{{2}}\lambda_{{3}} \left( {\lambda_{{3}}}^{2}-1 \right) }}
-
{\frac { d\lambda_2\wedge ds_1}{2i\pi s_{{1}}\lambda_{{2}}}}
-
{\frac {d\lambda_2\wedge ds_3}{2i\pi s_{{3}}\lambda_{{2}}}} 
-
{\frac {({\lambda_{{3}}}^{2}+1)d\lambda_3\wedge ds_1}{2i\pi s_{{1}} \left( {\lambda_{{3}}}^{2}-1 \right) \lambda_{{3}}}} 
- {\frac { d\lambda_3\wedge ds_3}{2i\pi s_{{3}}\lambda_{{3}}}}
+{\frac { ds_1\wedge ds_3}{2i\pi s_{{1}}s_{{3}}}}
\ee
and then a direct computation using the DeRham homotopy operator (after checking that the form above is indeed closed), we obtain 
\be
 \Thetavertex = 
{\frac { \left({
\lambda_{{3}}}^{2}\ln  \left( s_{{1}} s_3 \right) +\ln  \left( \frac{ s_{{1}}}{s_3}
 \right) 
  \right)\!\! { \d\lambda}_{{3}}}{2i\pi \lambda_{{3}} \left( {
\lambda_{{3}}}^{2}-1 \right) }}
-
{\frac {  \ln  \left( s_{{3}}   \right)\! {\d s}_{{1}}}{2i\pi s_{{1}}
}}
+
{\frac { \left(
\ln  \left( \frac{s_{{3}} s_1 (\l_3^2-1)}{\l_3}\right) 
  \right) \!\!\d\lambda_{{2}}}{2i\pi 
\lambda_{{2}}}}
-
{\frac { \left( 
{\lambda_{{1}}}^{2}\ln  \left(\frac{ s_3(\lambda_{{3}}^2-1 )\l_2}{s_1 \l_3}\right) 
+\ln  \left(\frac{s_3  s_{{1}} \l_2(\l_3^2-1)}{\l_3}\right) 
  \right)
\!\! { \d \lambda}_{{1
}}}{2i\pi  \left( {\lambda_{{1}}}^{2}-1 \right) \lambda_{{1}}}}
\nonumber
\ee
\ex

\subsection{Higher Poincar\'e\ rank singularities: the case of Painlev\'e\ II}

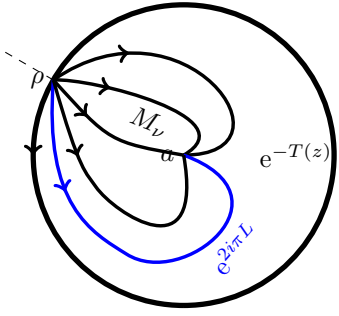
\begin{figure}
\begin{minipage}{0.3\textwidth}
\begin{center}
\begin{tikzpicture}[scale=2]
\draw [line width=2pt, postaction={decorate,decoration={{markings,mark=at position 0.75 with {\arrow[black,line width=1.5pt]{>}}}} }] (0,-1) arc [radius =1, start angle= -90, end angle= 270] ; 
\node [left] at (0:1.05) {${\rm e}^{-T(z)}$};
\draw [line width=1.3pt, postaction={decorate,decoration={{markings,mark=at position 0.75 with {\arrow[black,line width=1.5pt]{<}}}} }]
(0,0) 
.. controls (30: 0.5) and (140:0.8) ..  node [sloped,below] {$M_\nu $}(150:1);

\draw[dashed] (150:1)--(150:1.4);
\draw [line width=1.3pt, postaction={decorate,decoration={{markings,mark=at position 0.75 with {\arrow[black,line width=1.5pt]{<}}}} }]
(0,0) .. controls (0: 0.8) and (90:1.1) .. (150:1);
\node [left] at (0,0) {$a$};
\node [left] at  (150:1) {$\rho$} ;
\draw [blue, line width=1.3pt, postaction={decorate,decoration={{markings,mark=at position 0.75 with {\arrow[line width=1.5pt]{<}}}} }]
(0,0) .. node [sloped,below] {${\rm e}^{2i\pi L} $}
controls (-20: 0.8) and  (-80:0.9) .. (-120:0.76) .. 
controls (220:0.799) and (200:0.99) .. (150:1);

\draw [line width=1.3pt, postaction={decorate,decoration={{markings,mark=at position 0.75 with {\arrow[black,line width=1.5pt]{<}}}} }]
(0,0) .. 
controls (-80: 0.8) and (-140:0.6) .. (-160:0.62) .. 
controls (-170: 0.659) and (-170: 0.7).. (150:1);

\draw [line width=1.3pt, postaction={decorate,decoration={{markings,mark=at position 0.75 with {\arrow[black,line width=1.5pt]{<}}}} }]
(0,0) .. controls (170: 0.5)  .. (150:1);
\end{tikzpicture} 
\end{center}
\end{minipage}
\begin{minipage}{0.7\textwidth}
\caption{
The contours of the RHP within a ``toral circle'' (in the terminology of \cite{BertolaIsoTau}). In the figure,   {$M_\nu = (z-a)^L {\rm e}^{T(z)} S_\nu {\rm e}^{-T(z)} (z-a)^{-L} $}, the cut of the function $(z-a)^L$ is along the blue contour,  $T(z)$ is of the form $T(z) = \sum_{j=1}^{r+1} T_j (z-a)^{-j} + T_0$ and $T_0$ is a constant diagonal matrix chosen so that $(\rho-a)^L {\rm e}^{T(\rho)}= \1$, where $\rho$ is the point on the boundary of the toral circle where the various Stokes' contours meet.  }
\label{figStokes}
\end{minipage}
\end{figure}

If the system has also poles of higher order (under the same original genericity assumption  that the leading coefficient matrix of the singular part of the connection is semi-simple), then the corresponding RHP has additional contours of jumps to account for the Stokes' phenomenon. In view of the correction in Thm. \ref{thmreplace} we slightly modify their definition within the ``toral circle'' (Fig. 5 in \cite{BertolaIsoTau})  as in Fig. \ref{figStokes}. { We consider the case of only one singularity of higher Poincar\'e\ rank for clarity. } Supposing that $S_1,\dots S_{2r}$ are the Stokes' matrices and $L$ the diagonal matrix of the exponents of formal monodromy at a the  singularity, they must satisfy
\be
 S_1\cdots S_{2r} { \rm e}^{2i\pi L}=\1\label{Stokes}
\ee

\noindent and there is a contribution to $\Etavertex$ in the form (we denote $S_{2r+1}= {\rm e}^{2i\pi L}$)
\be
\Etavertex = \frac{-1}{4i\pi} \sum_{\ell=1}^{2r+1} \sum_{1\leq k<\ell} \tr \bigg(S_{[1:k-1]} \delta S_{k} S_{[k+1;\ell-1]} \wedge \delta S_{\ell} S_{[\ell+1:2r+1]}\bigg)
\ee
which is a closed two--form on the manifold \eqref{Stokes}. 
This type of contributions comes one for each higher Poincar\'e\ rank singularity. 
As an illustration, the example of Painlev\'e\ II
 is instructive. We follow the general formulation of (\cite{JMU2}, App.C); in this case we have six rays along $\varpi_\ell = {\rm e}^{ (\ell-1) i\pi/3 }\R_+$ and $\varpi_7 = {\rm e}^{-i\pi/ 6} \R_+$ with jumps 
\be
M_{1,3,5} = \le[\begin{array}{cc}
1 & s_{1,3,5}\\
0&1
\end{array}\ri]
\ ,\ \ \ 
M_{2,4,6} = \le[\begin{array}{cc}
1 & 0\\
s_{2,4,6} &1
\end{array}\ri]
\ ,\ \ M_7 =  \le[\begin{array}{cc}
\l & 0\\
0 &\l^{-1}
\end{array}\ri]
\ee
subject to the condition \eqref{localnomonodromy}.
We can solve for $s_2,s_4,s_6$ in  in the subset $\{s_1s_3s_5 \l\neq 0\}$ to give 
\be
 \left\{ s_{{2}}=-{\frac {s_{{5}}+s_{{3}}\lambda+\lambda\,s_{{1}}}{s_{
{3}}\lambda\,s_{{1}}}},s_{{4}}=-{\frac {\lambda\,s_{{1}}+s_{{5}}+s_{{3
}}}{s_{{3}}s_{{5}}}},s_{{6}}=-{\frac {s_{{5}}+s_{{3}}\lambda+s_{{1}}{
\lambda}^{2}}{s_{{1}}{\lambda}^{2}s_{{5}}}} \right\}
\ee
Then $\Etavertex$ becomes 
\be
\Etavertex ={\frac {\d s_1 \wedge \d\lambda }{2i\pi s_{{1}}\lambda}}
-
{\frac { \d s_3\wedge \d\lambda   }{2i\pi \lambda\,s_{{3}}}} 
+
{\frac {\d s_5\wedge  \d\lambda }{2i\pi \lambda\,s_{{5}}}}
-
{\frac {  \d s_3\wedge \d s_1}{2i\pi  s_{{3}}s_{{1}}}}
+
{\frac { \d s_5\wedge \d s_1 }{2i\pi s_{{1}}s_{{5}}}}
-
{\frac { \d s_5\wedge \d s_3}{2i\pi s_{{3}}s_{{5}}}}
\ee
Note that $\Etavertex$ defines a {\it symplectic form} on the Stokes' manifold, whereby all coordinates $\l,s_1,s_3,s_5$ are log-canonical and the form $\Thetavertex$ is 
\be
\Thetavertex =  
\ln  \left( {\frac {s_{{5}}s_{{1}}}{s_{{3}}}} \right) \frac{ \d\lambda}{4i\pi \l}
-
\ln  \left( {\frac {s_{{3}}\lambda}{s_{{5}}}} \right) \frac{ \d s_1}{4i\pi s_1}
+
\ln  \left( {\frac {s_{{1}}\lambda}{s_{{5}}}} \right)\frac{ \d s_3}{4i\pi s_3}
-
\ln  \left( {\frac {s_{{1}}\lambda}{s_{{3}}}} \right)\frac{ \d s_5}{4i\pi s_5}
\ee
The special case of PII as reported in \cite{BertolaIsoTau}  consists in 
\be
\left\{ \lambda=1,s_{{1}}=s_{{1}},s_{{2}}=-{\frac {s_{{3}}+s_{{1}}}{s
_{{1}}s_{{3}}+1}},s_{{3}}=s_{{3}},s_{{4}}=s_{{1}},s_{{5}}=-{\frac {s_{
{3}}+s_{{1}}}{s_{{1}}s_{{3}}+1}},s_{{6}}=s_{{3}} \right\} \\
\Etavertex = \frac { \d s_1\wedge \d s_3}{i\pi (s_1s_3+1)}\ ,\qquad
\Thetavertex = \frac{\ln(s_1s_3+1)}{2i\pi} \le(\frac{\d s_3}{s_3} - \frac {\d s_1}{s_1} \ri)
\ee

\section{Proof of Thm. \ref{thmreplace}}
\label{proof}
\bl
\label{lemmasame}
Let $\gamma$ be an oriented smooth arc without self-intersection  and 
let $\varphi: \gamma \times\gamma \to \C$ be a function which is locally analytic in each variable and such that $\varphi(z,w)= - \varphi(w,z)$. 
Then 
\be
\int_\gamma \ddz \int_\gamma \ddw \frac {\varphi(w,z)}{(w-z_-)^2}  = -\frac 1 2 \int_\gamma \ddz \pa_w \varphi(w,z)\bigg|_{w=z} .
\ee
\el
{\bf Proof.}    It is shown in Section 7  of \cite{Gakhov} that if $A(w,z)$ is H\"older (jointly) for $z,w\in \gamma$, then 
\be
\label{gakhov}
\int_\g \ddz \slint_\g\ddw \frac{A(w,z)}{(w-z)} = \int_\g \ddw \slint_\g\ddz\frac{A(w,z)}{(w-z)}.
\ee
where $\slint$ denotes the Cauchy's principal value integral.
We will use  Sokhotski-Plemelj's formula
\be
\int_\gamma \ddw \frac{ \Phi(w)}{w-z_\pm} = \slint_\gamma  \ddw\frac{ \Phi(w)}{w-z} \pm\frac 1 2 \Phi(z)\ ,\ \ \ z\in \gamma.
\label{sokho}
\ee
Since $\varphi (w,z)$ is (locally) jointly analytic in $z,w$, we can use Gakhov's result \eqref{gakhov} with $A(w,z) := \frac {\varphi(w,z)}{(w-z)}$, which is now also a jointly analytic function of its variables in a neighbourhood of $\gamma$. Note that $A(z,z) = \pa_w \varphi(w,z)|_{w=z}$ is well defined and $A(z,w)=A(w,z)$.
Now
\bea
 \int_{\gamma} \ddz \int_\gamma \ddw \le( \frac {\varphi(w,z)}{(w-z_-)^2}\ri) \mathop{=}^{\eqref{sokho}}  \int_{\gamma} \ddz \slint_\gamma \ddw \frac {A(w,z)}{(w-z)} -\frac 1 2 \int_\gamma \ddz  \pa_w \varphi(w,z)\bigg|_{w=z}.\label{13123}
 \eea
 We now show that the principal value integral is zero; to this end we use $A(z,w)=A(w,z)$, which holds for our case. Then 
 \bea 
\int_{\gamma} \ddz \slint_\gamma \ddw \frac {A(w,z)}{(w-z)} 
 = \int_{\gamma} \ddz \slint_\gamma \ddw \frac {A(z,w)}{(w-z)}  \mathop{=}^{\eqref{gakhov}} \int_{\gamma} \ddw \slint_\gamma \ddz \frac {A(z,w)}{(w-z)} \mathop{=}^{z\leftrightarrow w}-\int_{\gamma} \ddz \slint_\gamma \ddw \frac {A(w,z)}{(w-z)}
\eea
Thus the contribution of the principal value integral in \eqref{13123} is zero.\QED

{\bf Proof of Thm. \ref{thmreplace}}
We shall denote by ${\Sigma\gamma} _\epsilon = \Sigma\gamma \setminus \bigcup_{v\in\mathfrak V} \D_\epsilon^{(v)} =: \Sigma\gamma \setminus \D_\epsilon $  the support of the jumps minus small $\epsilon$--disks around the point of self-intersection of ${\Sigma\gamma}$; we use the notation  
\be
\label{defphi}
\varphi (w,z):= \tr\bigg(\G_-(w) \Xi(w) \G_-^{-1}(w)\wedge  \G_-(z) \Xi (z) \G_-^{-1}(z) \bigg)\, \ \ \Rightarrow \ \       \ \varphi(z,w) = -\varphi(w,z).
\ee
 Lemma 2.1 in \cite{BertolaIsoTau} yields the formula (in the new notation)
\be
\delta \le(\G_-^{-1}(z) \G_-'(z)\ri) = \int_{\Sigma\gamma}\ddw  \frac {\G_-^{-1}(z)\G_-(w) \Xi(w) \G_-^{-1}(w) \G_-(z)}{(z_--w)^2}
\ee
whence we compute the exterior derivative as follows
\bea
\delta  \omega_{_{M}}  = \delta  \int_{{\Sigma\gamma} } \ddz \tr\le(\Gamma_-^{-1}(z) \G_-'(z) \Xi (z)\ri) = 
 \\
 =
  \int_{{\Sigma\gamma} } \ddz \int_{\Sigma\gamma}  \ddw \tr\le( \frac {\Gamma_-^{-1}(z)\Gamma_-(w) \Xi (w)  \Gamma_-^{-1}(w)\wedge \Gamma_-(z) \Xi(z)}{(z_--w)^2}\ri) 
 +
 \int_{{\Sigma\gamma} } \ddz \tr\le(\Gamma_-^{-1}(z) \G_-'(z)  \delta  \Xi (z)\ri)
\eea
Note that $\Xi = \delta M M^{-1}$ satisfies $\delta \Xi =\Xi \wedge  \Xi $ (i.e. $\delta \Xi\lfloor_{\pa,\wt \pa}  =  \pa \Xi_{\wt \pa} -\wt \pa \Xi_{\pa}  = [\Xi_\pa, \Xi_{\wt \pa}]$)
and hence (using the cyclicity of the trace)  
\be
\label{deltaomega}
\delta  \omega_{_{M}}  =   \int_{{\Sigma\gamma} } \ddz \int_{\Sigma\gamma}  \ddw 
\frac{ \varphi(w,z)}{(z_--w)^2} 
 +
 \int_{{\Sigma\gamma} } \ddz \tr\le(\Gamma_-^{-1}(z) \G_-'(z) \Xi(z) \wedge \Xi (z)  \ri)
\ee
The issue is the computation of the iterated integral; as an iterated integral it is convergent but its value depends on the order of integration.
To see why  it is convergent,  we need to make sure that the inner integral does not have too severe  singularities; these may occur at the intersection points of the arcs because the derivative of the Cauchy transform may have poles. 
{ For an ``inessential'' intersection point (where the jump matrices tend exponentially to the identity) we can excise a small disk and easily evaluate the contribution to be infinitesimal as the radius tends to zero (this procedure will be considered in more detail later on). So let us consider the issue of local convergence near an ``essential" intersection point $v\in \mathfrak V$.}
On the incident arcs $\g_1,\dots, \g_n$ (in counterclockwise order)  define 
\be
\G_j(z) := \G_-(z)\bigg|_{z\in \g_j}\ ,\ \ \varphi_{k j }(w,z) = \varphi(w,z) \bigg|_{z\in \g_j\atop w\in \g_k}\ \ \hbox{ etc.}
\ee
Without loss of generality we can assume that all arcs $\g_j$ are oriented away from $v$. 
If a ray is incident, then this requires us to locally re-define $\G_j \mapsto \G_j M_j$, $M_j \mapsto M_j^{-1}$ and reverse the orientation in the integral, thanks to the obvious formula
\bea
 \G_k\pa M_k M_k^{-1} \G^{-1}_k  =- (\G_kM_k)\pa (M_k^{-1}) M_k(\G_kM_k)^{-1} 
\eea
(note that $\G_- M = \G_+$ becomes the $-$ boundary value in the reversed orientation).

Under Assumption \eqref{localnomonodromy}, we can locally express the analytic extensions of each $\G_j(z)$ in terms of the analytic extension of $\G_1(z)$ to a full neighbourhood of $v\in \mathfrak V$;
\be
\G_k (z)= \G_\ell (z) M_{[\ell:k-1]}(z) = \G_1(z)M_{[1:k-1]}(z)\ ,\ \ \ell<k.\label{Gammaellk}
\ee
where $M_{[a:b]}(z) = M_a (z)M_{a+1} (z)\cdots M_b(z)$.
Taking the differential $\delta$ of  the local condition  \eqref{localnomonodromy} we find  (evaluation being understood at $z=v$) 
\bea
\sum_{j=1}^n M_{[1:j-1]} \Xi_{j} M_{[j:n]}  =\sum_{j=1}^n \G_1^{-1} \G_j \Xi_{j}\G_j^{-1} \G_1   \equiv 0 \ \ \Rightarrow\ \ 
\sum_{j=1}^n  \G_j \Xi_{j}\G_j^{-1}   \equiv 0\label{sum0}\\
\Rightarrow \ \ \forall \ell = 1,\dots, n\ \ \ \ \sum_{k=1}^n \varphi_{k\ell}(v,w) = \sum_{ k=1}^n \varphi_{\ell k} (z,v) \equiv 0 \label{phireg}
\eea
Now, consider the part of the  inner integral along $\g_\ell$ for $z\not\in {\Sigma\gamma}$ near $v$; since $\varphi_{\ell,m}(z,w)$ extends to a locally analytic function in the neighborhood of $v$, the properties of the Cauchy transform immediately imply the following local identity of analytic functions
\be
J_{\ell m}(z) = \int_{\g_\ell} \frac {\varphi_{\ell m}(w,z)}{(w-z)^2}\ddw  = \frac {\varphi_{ \ell m}(v,v)} {z-v} + \ln_{\ell} (z-v) F_{\ell m}(z) + G_{\ell m}(z).
\label{Jmell}
\ee
Here $F_{\ell m}, G_{ \ell m}$ are locally analytic functions and $\ln_\ell(z-v)$ stands for the logarithm with the branch-cut extending along $\gamma_\ell$. Now we see that the total integral over $\Sigma\gamma$ involves summing over the incident arcs at $v$ and then \eqref{phireg} implies that the pole in \eqref{Jmell} cancels out in the summation so that  the integral is locally convergent in the ordinary sense (irrespectively of the boundary values of the logarithms).
Having established the convergence of the integral, we now choose  $\epsilon$ sufficiently small so that the various disks $\D_\epsilon^{(v)}$ are disjoint. We then have
\bea
\int_{{\Sigma\gamma} } \ddz&\&  \int_{\Sigma\gamma}  \ddw 
\frac{ \varphi(w,z)}{(z_--w)^2} = \underbrace{\int_{{\Sigma\gamma}_\epsilon } \ddz \int_{{\Sigma\gamma}_\epsilon}  \ddw 
\frac{ \varphi(w, z)}{(z_--w)^2} }_{A_\epsilon} 
+\\
+&\&
\underbrace{\le(\int_{{\Sigma\gamma}_\epsilon } \ddz \int_{{\Sigma\gamma}\cap \D_\epsilon}  \ddw  + \int_{{\Sigma\gamma}\cap \D_\epsilon}  \ddz  \int_{{\Sigma\gamma}_\epsilon }\ddw \ri)
\frac{ \varphi(w, z)}{(z -w)^2} }_{B_\epsilon} 
+
\underbrace{ \int_{{\Sigma\gamma}\cap \D_\epsilon } \ddz \int_{{\Sigma\gamma}\cap \D_\epsilon}  \ddw 
\frac{ \varphi(w,z)}{(z_--w)^2} }_{C_\epsilon}
\nonumber 
\eea
The expression $A_\epsilon$ consists only of integrations over  non-intersecting arcs and thus we can apply Lemma \ref{lemmasame}
\be
A_\epsilon  = -\frac 1 2\int_{{\Sigma\gamma}_\epsilon} \ddz \pa_w \varphi(w,z) \bigg|_{w=z}
\ee
This term clearly admits a limit as $\epsilon=0$ equal to $A_0$; a short computation gives 
\be
\pa_w\varphi(w,z)\bigg|_{w=z} \mathop{=}^{\eqref{defphi}} 
 2 \tr\bigg( \G_-^{-1}(z) \G_-'(z)\Xi(z) \wedge \Xi(z) \bigg)   + \tr\bigg(\Xi'(z) \wedge \Xi(z)\bigg) 
\ee 
and hence 
\be
\label{A0} 
A_0 = -\int_{\Sigma\gamma} \ddz \le\{\tr\bigg( \G_-^{-1}(z) \G_-'(z)\Xi(z) \wedge \Xi(z) \bigg) + \frac 1 2 \tr\bigg(\Xi'(z) \wedge \Xi(z)\bigg) \ri\}
\ee

The remaining issue is the evaluation of $B_\epsilon, C_\epsilon$: as for $B_\epsilon$ we now show that it is identically zero.
The inner integration in $w$ and the outer integration in $z$ have common points at ${\Sigma\gamma} \cap \pa \D_\epsilon$; let $c$ be one of these points. 
 \\
The integrand in $B_\epsilon$ is actually $L^1$ integrable over  $\Sigma\gamma_\epsilon \times (\Sigma\gamma\cap \D_\epsilon)$ (and the reversed) because near the points common to those two sets (on the boundary of $\D_\epsilon$) the behaviour of the integrand is 
\be
\frac{\varphi(w,z)}{(z-w)^2} = \frac {C}{(z-w)} + \mathcal O(1)
\ee

\noindent and hence the local  nature  of the integral is the same as  the convergent integral $\int_0^\epsilon \int_{-\epsilon}^0 \frac {\d x \d y}{x+y}$.
Thus the interchange of order of  integral is allowed by Fubini's theorem and we conclude that $B_\epsilon \equiv 0$ (using $\varphi(z,w) = -\varphi(w,z)$) for all $\epsilon$ (sufficiently small). 

It remains to analyze the term $C_\epsilon$; it is clear (due to the skew-symmetry of $\varphi$) that the only contributions to the double integral may come from $(z,w)$ in a neighborhood of the {\it same} vertex $v\in \mathfrak V$. { Moreover, a simple estimate shows that if $v$ is an ``inessential" vertex, then the contribution tends to zero as $\epsilon \to 0$. For this reason we focus below only on the ``essential'' vertices.}

Consider one of them and denote the incident arcs in $\Sigma\gamma \cap \{|z-v|<\epsilon\} =  \bigcup_{\ell=1}^{n_v}  \g_\ell^{(v)}$ by 
$\gamma_\ell, \ \ell =1,\dots, n$; denote by $\sigma_\ell$ the distal endpoints of the arcs (at distance $\epsilon$ from $v$). We denote with $\varphi_{\ell m}^{(v)}  =\ds  \lim_{z, w \to v\atop z\in \g_\ell, w\in \g_m }\varphi_{\ell m}(z,w)$ for brevity (we also omit the superscript $^{(v)}$ since we consider one vertex at a time).  Then 
\be
\int_{\Sigma\gamma \cap \D_\epsilon ^{(v)}}\ddz\int_{\Sigma\gamma \cap \D_\epsilon ^{(v)}}\ddw \frac {\varphi(w,z)}{(z_--w)^2} = 
\sum_{\ell=1}^{n} \sum_{m=1}^n \int_{\g_\ell}\ddz \int_{\g_m}\ddw \frac {\varphi_{ m \ell }(w,z)}{(z_--w)^2}
\ee
The term with $\ell=m$ yields a convergent integral that is handled by Lemma \ref{lemmasame} and tends to zero as $\epsilon\searrow 0$ since the length of $\gamma_\ell$ is $\mathcal O(\epsilon)$ (and the integrand is bounded).  Let us now consider the remaining terms;
\bea
\mathbf J &\&:= 
\sum_\ell \int_{\g_\ell} \ddz \sum_{m\neq \ell} \int_{\g_m} \ddw
 \frac {\varphi_{ m \ell }(w,z)}{(w-z)^2} =
\cr
&\&=\underbrace{\sum_\ell \int_{\g_\ell} \ddz \sum_{m\neq \ell} \int_{\g_m} \ddw \frac {\varphi_{ m \ell }(w,z) - \varphi_{m \ell }}{(w-z)^2}}_{(\star)}
+
\sum_\ell \int_{\g_\ell} \ddz \sum_{m\neq \ell} \int_{\g_m} \ddw \frac {\varphi_{m \ell }}{(w-z)^2} 
\eea
Each double  integral in the sum marked $(\star)$ is a regularly convergent integral  because of the regularization constant that we have added and subtracted; the other integral, on the contrary, is a singular integral and it depends on the order of integration. On the integral $(\star)$  we can swap order of integration and relabel $z\leftrightarrow w, \ell\leftrightarrow m$, and then use the skew symmetry $\varphi_{\ell m}(z,w) = -\varphi_{m \ell} (w,z)$,  like so
\bea
\mathbf J&\&=
\sum_m \int_{\g_m} \ddw \sum_{\ell \neq m} \int_{\g_\ell } \ddz \frac {\varphi_{m\ell }(w,z) -\varphi_{m\ell }}{(w-z)^2}
+
\sum_\ell \int_{\g_\ell} \ddz \sum_{m\neq \ell} \int_{\g_m} \ddw \frac {\varphi_{m \ell }}{(w-z)^2} 
\mathop{=}^{z\leftrightarrow w \atop \ell\leftrightarrow m}
\\
&\&=
\sum_\ell \int_{\g_\ell} \ddz \sum_{m \neq \ell} \int_{\g_m } \ddw \frac {\varphi_{\ell m }(z,w) -\varphi_{\ell m }}{(z-w)^2}
+
\sum_\ell \int_{\g_\ell} \ddz \sum_{m\neq \ell} \int_{\g_m} \ddw \frac {\varphi_{m \ell }}{(w-z)^2} .
\label{andrei}
\eea
{  In the step above we only have re-labeled the dummy variables of integration. Now we can use the skew--symmetry $\varphi_{\ell m }(z,w) =-\varphi_{m\ell }(w,z)$ and $\varphi_{\ell m}=-\varphi_{m\ell}$ in the first integral. The term containing $\varphi_{\ell m}(z,w)$ yields back $-\mathbf J$ while the term containing $\varphi_{m\ell}$ adds to the last integral in \eqref{andrei}. Thus we continue the chain of equalities:}\\
\bea
&\&=
-\mathbf J  
+2
\sum_\ell \int_{\g_\ell} \ddz \sum_{m\neq \ell} \int_{\g_m} \ddw \frac {\varphi_{m\ell}}{(w-z)^2} 
=
-\mathbf J + 2 \sum_\ell \int_{\g_\ell} \ddz \sum_{m\neq \ell} \frac {\varphi_{  m \ell  }}{2i\pi(z-{\sigma}_m)}
\eea
\begin{figure}
\begin{minipage}{0.49\textwidth}
\begin{center}
\begin{tikzpicture}[scale=1.4]
\draw [fill=gray!10!white] circle [radius =1];
\node at (-0.5,-0.5){$\D_\epsilon$};
\draw[dashed]
(10:1)--(10:1.4);
\draw[postaction={decorate,decoration={{markings,mark=at position 0.75 with {\arrow[black,line width=1.5pt]{>}}}} }]
(10:0)--(10:1);
\node at (0,-0.20){$v$};
\node at (-10:0.6){$\g_1$};
\node at (55:0.6){$\g_2$};
\node at (150:0.6){$\g_\ell$};
\node[below] at (10:1.2){$\sigma_1$};
\node [above] at (30:1.2){$\sigma_2$};
\node [above] at (170:1.2){$\sigma_\ell$};
\draw[dashed]
(30:1)--(30:1.4);
\draw[postaction={decorate,decoration={{markings,mark=at position 0.75 with {\arrow[black,line width=1.5pt]{>}}}} }]
(0,0)--(30:1);
\draw[dashed]
(170:1)--( 170:1.4);
\draw[postaction={decorate,decoration={{markings,mark=at position 0.75 with {\arrow[black,line width=1.5pt]{>}}}} }]
(170:0)--( 170:1);
\draw [fill] (170:1) circle[radius=0.03];
\draw [fill] (10:1) circle[radius=0.03];
\draw [fill] (30:1) circle[radius=0.03];
\end{tikzpicture} 
\end{center}
\vspace{-10pt}
\captionof{figure}{Arcs near a vertex.}\label{figsigma}
\end{minipage}
\begin{minipage}{0.5\textwidth}
\begin{center}
\begin{tikzpicture}[scale=1.3]
\draw[postaction={decorate,decoration={{markings,mark=at position 0.75 with {\arrow[black,line width=1.5pt]{>}}}} }]
(0,0)--(10:2);
\draw[green](0,0)--(0:2);
\node at (5:2){$\bullet \, c$};
\node at (15:1.8){$\g_1$};
\node at (0,-0.20){$v$};
\node at (35:1.8){$\g_2$};
\node at (160:1.4){$\g_\ell$};
\node at (355:1.8){$\green{\g_0}$};
\node at (335:1.8){${\g_n}$};
\node at (180:1.4){$z$};
\draw[postaction={decorate,decoration={{markings,mark=at position 0.75 with {\arrow[black,line width=1.5pt]{>}}}} }]
(0,0)--(30:2);
\draw[postaction={decorate,decoration={{markings,mark=at position 0.75 with {\arrow[black,line width=1.5pt]{>}}}} }]
(0,0)--(340:2);
\draw[postaction={decorate,decoration={{markings,mark=at position 0.75 with {\arrow[black,line width=1.5pt]{>}}}} }]
(0,0)--( 170:2);
\end{tikzpicture}
\end{center}
\vspace{-18pt}
\captionof{figure}{Vertex contribution to the exterior derivative.}
\label{figc}
\end{minipage} 
\end{figure}

Here ${\sigma}_m =  \gamma_m \cap \D_\epsilon$ (Fig. \ref{figsigma}). Solving for $\mathbf J$ we obtain finally:
\be
\mathbf J = \sum_\ell \int_{\g_\ell} \ddz \sum_{m\neq \ell} \frac {\varphi_{ m \ell }}{2i\pi(z-{\sigma}_m)}
=\sum_\ell \sum_{m\neq \ell}\frac{\varphi_{m \ell }}{(2i\pi)^2} \ln_\ell \le(\frac{{\sigma}_\ell - {\sigma}_m} {v-{\sigma}_m}\ri)
\eea
To compute the last expression we proceed as follows; first we observe that it is independent of $v $ and ${\sigma}_\ell$'s. Indeed, differentiating we get
\be
\pa_v \mathbf J &\&= \sum_\ell \sum_{m\neq \ell}\frac{\varphi_{m \ell }}{(2i\pi)^2} \frac 1{v-{\sigma}_m} \mathop{=}^{\eqref{phireg} + (\varphi_{\ell \ell}=0)} 0
\cr
\pa_{{\sigma}_j}  \mathbf J&\& = \sum_{m\neq j} \frac {\varphi_{ m j }}{(2i\pi)^2}  \frac 1{{\sigma}_j - {\sigma}_m} 
+  \sum_{\ell \neq j} \frac {\varphi_{j \ell }}{(2i\pi)^2}  \le(\frac 1{{\sigma}_j - {\sigma}_\ell} - \frac 1{v-{\sigma}_j}\ri) \mathop{=}^{ \eqref{phireg}}
\cr
&\&=\sum_{m\neq j} \frac {\varphi_{ m j}}{(2i\pi)^2}  \frac 1{{\sigma}_j - {\sigma}_m} 
+  \sum_{m \neq j} \frac {\varphi_{j m }}{(2i\pi)^2}  \le(\frac 1{{\sigma}_j - {\sigma}_m}\ri)  \mathop{=}^{\varphi_{m \ell  } = -\varphi_{\ell m}} 0
\ee
Thus we can compute $\mathbf J$ by arranging ${\sigma}_\ell$'s as we wish; to do so, we re-write it back as a double integral 
\bea
\mathbf J  &\&=  \sum_\ell \int_{\g_\ell} \ddz \sum_{m\neq \ell} \int_{\g_m}\ddw \frac {{\varphi_{m \ell}}}{(z-w)^2}
=
\cr
&\&= \sum_\ell \int_{\g_\ell} \ddz \sum_{m\neq \ell} \int_{\g_m}\ddw \frac {{\varphi_{m \ell}}\le(1 - \frac {v-c}{2(w-c)} - \frac {v-c}{2(z-c)}\ri)}{(z-w)^2}
+
\frac 1 2 \sum_\ell \int_{\g_\ell} \ddz \sum_{m\neq \ell} {\varphi_{m \ell}}\int_{\g_m}\ddw \frac { \frac {v-c}{w-c} + \frac {v-c}{z-c}}{(z-w)^2}
\eea
The point $c$ is an arbitrary point not on any of the segments.
The first integral is a regular convergent integral (the integrand is in $L^1$ near $z=w=v$) and exchanging the order of integration and then renaming the variables yields the same expression with a minus sign: hence we conclude that it is zero. We are thus left with the second term, which we now know is also independent of $\sigma_\ell$'s. To compute it more conveniently,  we send all ${\sigma}_\ell$'s to infinity along distinct directions: the value of the expression, as we know, is independent of ${\sigma}$'s,  so that now the arcs $\gamma_\ell$ are simply pairwise distinct  rays issuing from $v$. 
In the limit we obtain the following
\be
\mathbf J =\underbrace{ \frac 1 2 \sum_\ell \int_{\g_\ell} \ddz \sum_{m\neq \ell} {\varphi_{m \ell}}\int_{\g_m}\ddw \frac {  \frac {v-c}{z-c}}{(z-w)^2}}_{(\dagger)}
+\underbrace{\frac 1 2 \sum_\ell \int_{\g_\ell} \ddz \sum_{m\neq \ell} {\varphi_{m \ell}}\int_{\g_m}\ddw \frac { \frac {v-c}{w-c}}{(z-w)^2}}_{(\heartsuit)}
\label{134}
\ee
The term $(\dagger)$ is zero: indeed it is 
\be
(\dagger) = \frac 1 2 \sum_\ell \int_{\g_\ell} \ddz \sum_{m\neq \ell} {\varphi_{m \ell}} \frac {  \frac {v-c}{z-c}}{2i\pi(z-v)} \mathop{=}^{\eqref{phireg}}0.
\ee
In the remaining term { $(\heartsuit)$} , we write the { inner integral}  in partial fractions
\be
\int_{\g_m} \ddw \frac {\frac {v-c}{w-c}}{(w-z)^2} = \frac{v-c}{(z-c)^2} \int_{\g_m} \ddw
\le( \frac 1{w-c} - \frac 1 {w-z} \ri) + \frac {c-v}{c-z} \underbrace{\int_{\g_m}\ddw \frac 1{(w-z)^2}}_{=-\frac 1 {z-v} }\ \  z\in \g_\ell
\label{11222}
\ee
The first integral depends on $m$; to make the dependence manifest, we choose $c$ on the right of $\g_1$ but to the left of $\g_n$  (the final result is independent of this choice) and rotate the contour of integration in \eqref{11222} {counter}clockwise to a direction  between $c$ and $\g_n$ (denoted $\g_0$, see Fig. \ref{figc}): \\
\begin{minipage}{0.699\textwidth}
\be
&\& \frac{v-c}{(z-c)^2} \int_{\g_m} \ddw
\le( \frac 1{w-c} - \frac 1 {w-z} \ri) =\cr
&\& = \frac{v-c}{(z-c)^2} \int_{\g_0} \ddw
\le( \frac 1{w-c} - \frac 1 {w-z} \ri) -\frac {v-c}{(z-c)^2} \le\{
\begin{array}{cc}
1 & m<\ell\\
0 & m>\ell
\end{array}
\ri.
\ee
\end{minipage}
\\[5pt]
{ The last term is due to the residue that we picked up while rotating $\gamma_m$ counterclockwise, and the fact that $z\in \gamma_{\ell}$}.
Thus, summarizing  
\be
\int_{\g_m} \ddw \frac {\frac {v-c}{w-c}}{(w-z)^2} = \frac {v-c}{(z-c)^2} \ln_0\le(\frac { v-z}{v-c} \ri)
+ \frac {c-v}{(z-c)(z-v)}
-\frac {v-c}{(z-c)^2} \le\{
\begin{array}{cc}
1 & m<\ell\\
0 & m>\ell
\end{array}
\ri.
\ee
The first two terms are  independent of $m$ and hence they give a zero contribution to the term $(\heartsuit)$ in  \eqref{134} because of the condition \eqref{phireg}. We are thus left with 
\be
\mathbf J =-\frac 1 2 \sum_\ell \sum_{m< \ell} {\varphi_{m \ell}}\int_{\g_\ell} \ddz 
\frac {v-c}{(z-c)^2} = -\frac 1 {4i\pi} \sum_\ell \sum_{m< \ell} {\varphi_{m \ell}}
\ee   
Finally, one reads off the definition of $\varphi_{m\ell}$,
\be
\varphi_{m\ell}  = \lim_{z,w\to v\atop z\in \g_m, w\in\g_\ell}\tr\le( \Gamma_m\Xi_m \Gamma_m^{-1}\wedge  \Gamma_\ell\Xi_\ell \Gamma_\ell^{-1} \ri)
\ee
and  using \eqref{Gammaellk} we obtain 
the terms in the sum appearing in \eqref{corr21}. 
\QED

\def\cprime{$'$}

\end{document}